\documentclass[aps,prb,reprint,showpacs,superscriptaddress]{revtex4-2}
\usepackage{graphicx}
\usepackage{dcolumn}
\usepackage{bm}
\usepackage{amssymb}
\usepackage{amsmath}
\usepackage{subfigure}
\usepackage{epstopdf}
\usepackage{float}
\usepackage{mathrsfs}
\usepackage{xcolor}

\begin{document}
\title{Driven-dissipative Quantum Dynamics in Cavity Magnon-Polariton System}
\author{Guogan Zhao}
\affiliation{School of Physics, Nankai University, Tianjin 300071, China}
\author{Yong Wang}
\email[]{yongwang@nankai.edu.cn}
\affiliation{School of Physics, Nankai University, Tianjin 300071, China}
\author{X.-F. Qian}
\affiliation{Center for Quantum Science and Engineering, and Department of Physics, Stevens Institute of Technology, Hoboken, New Jersey 07030, USA}

\begin{abstract} 
The dynamics of arbitrary-order quantum correlations in a cavity magnon-polariton system are investigated based on the quantum master equation in the coherent state representation.  The phenomena of Rabi-like oscillation and level repulsion of the average cavity-photon number agree remarkably well with existing experimental observations.  The competing nature of coherent and incoherent components in these two cases is further revealed by the second-order quantum coherence of the cavity photons and magnons,  which can be systematically tuned by the driving microwave and thermal bath.   Our results demonstrate the rich higher-order quantum dynamics induced by magnetic light-matter interaction,  and serve as an indispensable step toward exploring nonclassical states for cavity photons and magnons in quantum cavity magnonics. 
\end{abstract}
\maketitle

\section{Introduction} 
The successful realization of strong coupling between photons and magnons in microwave cavities~\cite{Imamoglu2009, Flatte2010-1, Flatte2010-2, Huebl2013, Nakamura2014,Tang2014,Tobar2014,Hu2015-1} has brought a new member into the family of cavity quantum electrodynamics (QED) systems \cite{Review}.  Experimental measurements of this hybrid quantum system have revealed the formation of magnon-polariton quasiparticles,  which can be tuned using bias magnetic fields \cite{Huebl2013, Nakamura2014,Tang2014,Tobar2014,Hu2015-1,Haigh2015}, the cavity configuration \cite{Hu2015-2,Hu2018-NC,APL2020},  a DC voltage \cite{Hu2016-APL},  the experimental temperature~\cite{Tobar2018,Boventer2018},  Floquet engineering~\cite{XFZhang2020},  and so on.  Extended studies have also found that the intrinsic nonlinearity of magnon-magnon interactions can lead to bistable behaviors of cavity magnon-polaritons \cite{You2018,Hu2018-PRB},  and that dissipative magnon-photon coupling will result in level attraction~\cite{Hu2018-PRL,Xia2018,Xiao2019} and non-Hermitian physics~\cite{You2018-NC,Yan2019}.  Moreover, the cavity magnon-polariton has been utilized to develop gradient memory~\cite{Tang2015} and logic devices~\cite{Hu2019},  to manipulate spin currents~\cite{Bai2017} and magnons\cite{WuY2019,LiF2020},  and to generate quantum entanglement \cite{Li2018,Li2020,Yuan2020,Yuan2020-PRB,Qian2021} or Schr\"{o}dinger's cat states \cite{Sharma2021,QYHe2021}. Coherent control of the dynamics of cavity magnon-polaritons has also been experimentally demonstrated \cite{Ruoso2017,Weides2020}, paving the way to the realization of universal information processing.

Current experimental observations of dynamical features in cavity magnon-polariton systems have mostly focused on probing the power spectrum of the reflected or transmitted microwave field~\cite{Huebl2013,Nakamura2014,Tang2014,Tobar2014,Hu2015-1,Review,Haigh2015,Hu2015-2,Hu2018-NC,APL2020,Hu2016-APL,Tobar2018,Boventer2018,XFZhang2020,You2018,Hu2018-PRB,Hu2018-PRL,You2018-NC,Tang2015,Hu2019,Ruoso2017,Weides2020},  which depends on the average number of microwave photons in the cavity.  However,  as pointed out initially by Glauber~\cite{GlauberPRL,Glauber1963-1,Glauber1963-2} after the seminal Hanbury Brown-Twiss experiment \cite{HBT},  infinite sets of field correlation functions are necessary in order to fully characterize the quantum statistical properties of electromagnetic fields \cite{GlauberPRL,Glauber1963-1,Glauber1963-2}.  Therefore,  in addition to the average photon number,  which is directly related to the first-order field correlation,  higher-order field correlations are also crucial components in quantum optics \cite{QuanOpt1,QuanOpt2,QuanOpt3}.  For example,  the bunching and antibunching phenomena of photons,  which have been observed in classical and nonclassical optical fields, respectively,  are relevant to the second-order field correlation \cite{QuanOpt1,QuanOpt2,QuanOpt3}.  Furthermore,  higher-order correlations have also played an essential role in many other quantum systems,  such as circuit QED systems \cite{Rebic2009,Bozyigit2011,LangC2011,EichlerC2012,LangC2013, Peng2016,Gasparinetti2017,Rolland2019},  cavity exciton-polariton systems \cite{Higher1,PRL2018,NatMat2019},  cavity optomechanics \cite{OptoMech1,OptoMech2},  ultracold atoms \cite{Atom1,Atom2,Atom3,Atom4,Atom5,Atom6,Higher3,Atom7,Atom8,Atom9},  and metal-magnet hybrid structures \cite{SpinCurr}.   Although the second-order quantum coherence has been utilized to characterize the magnon blockade effect in magnon-qubit systems recently\cite{WuY2019,LiF2020},   a systematic investigation of higher-order quantum correlations in the widely observed dynamical processes in cavity magnon-polariton systems is still missing,  which seriously hinders the further developments of cavity magnonics beyond the semiclassical level\cite{Review}.  

In this work,  the driven-dissipative dynamics in a cavity magnon-polariton system has been thoroughly studied based on the quantum master equation in the coherent state representation.  In Sec.  II,  a Fokker-Planck equation of quasiprobability distribution function and a group of hierarchical equations of arbitrary-order correlation functions have been established for the coupled cavity photons and magnons.  Then the theoretical approach has been applied to investigate the average number of cavity photons and magnons in Sec. III.  A and their second-order quantum coherence in Sec.  III.  B for two typical experimental scenarios.  Possible experimental techniques to measure higher-order correlation functions of cavity photons are briefly discussed in Sec.  III.C.   Finally,   the results are concluded in Sec.  IV.  

\begin{figure}[!ht]
  \centering
  \includegraphics[width=0.5\textwidth,clip]{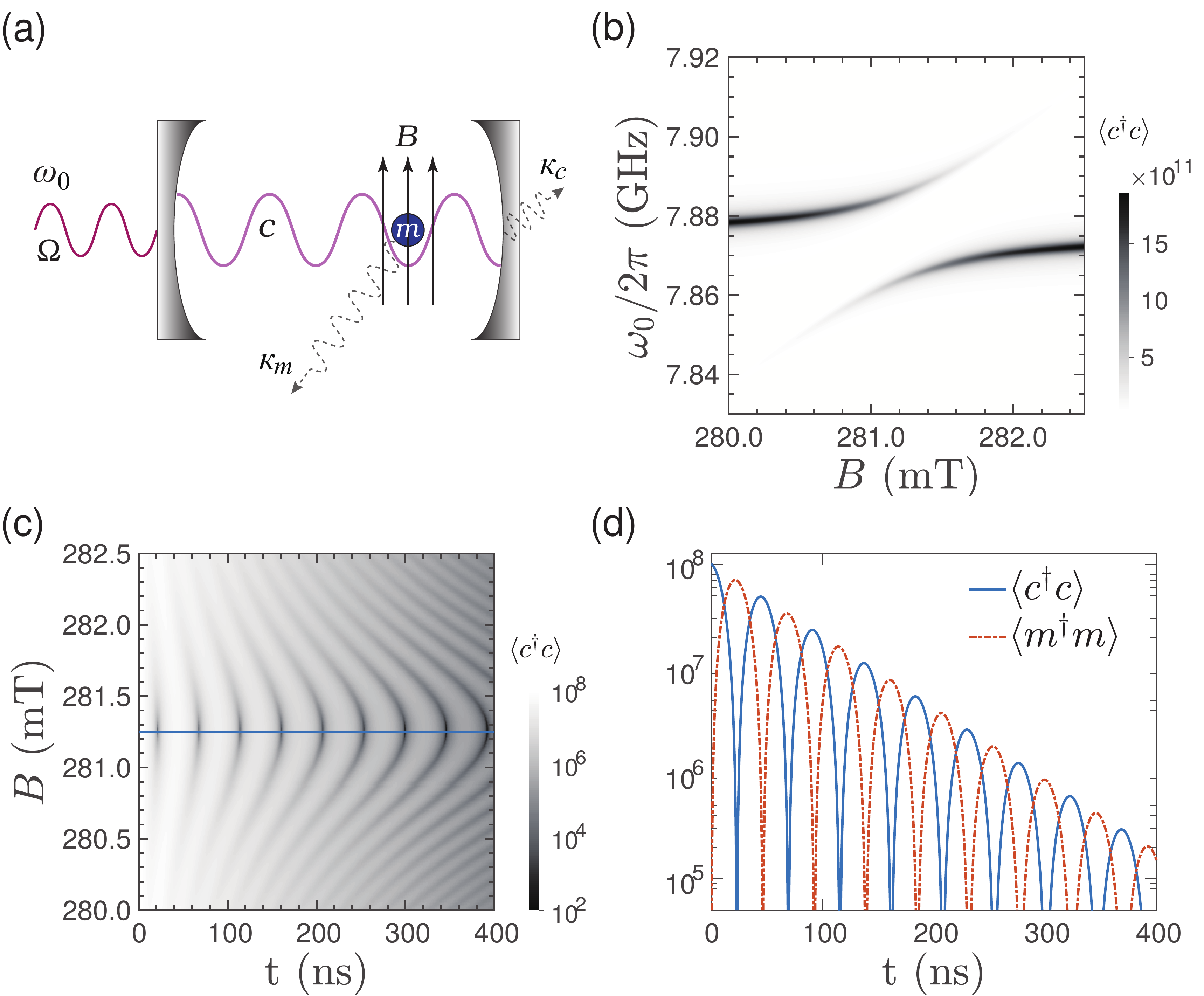}
  \caption{(Color online) (a) A schematic diagram of the cavity magnon-polariton system.  A magnet denoted by $m$ is coupled to the cavity microwave field $c$.  A bias magnetic field $B$ is applied to tune the magnon frequency $\omega_{m}=\gamma B$ with gyromagnetic ratio $\gamma$.  $\kappa_{c}$ and $\kappa_{m}$ are the damping rates for the cavity photons and magnons, respectively.  An external microwave field with strength $\Omega$ and frequency $\omega_{0}$ can be applied to continuously drive the cavity.  (b) The average cavity-photon number $\langle c^{\dag}c\rangle$ as a function of the bias magnetic field $B$ and driving frequency $\omega_{0}$ in the continuous drive scenario,  with driving strength $\Omega/2\pi=2\times 10^{12}$~Hz.  (c) The time-evolution of the average cavity-photon number $\langle c^{\dag}c\rangle$ for different bias magnetic fields $B$ after injecting $10^{8}$ coherent microwave photons in an initial pulse.  (d) The Rabi-like oscillation of the average number of cavity photons (blue solid line) and magnons (red dashed line) with a zero-detuned bias magnetic field $B=281.25$~mT,  i.e., $\omega_{m}=\omega_{c}$.   The other simulation parameters in (b), (c), and (d) are set to \cite{Tang2014}: $\omega_{c}/2\pi=7.875$~GHz,  $\kappa_{c}/2\pi=1.35$~MHz,  $\kappa_{m}/2\pi=1.06$~MHz,  $g/2\pi=10.8$~MHz,  and $T=300$~K.} 
 \label{Fig1}
\end{figure}

\section{Theoretical Model and Quantum Dynamical Equations} The cavity magnon-polariton system under consideration is schematically illustrated in Fig.~1(a).  A highly-polished YIG sphere with a diameter of $0.36$~mm,  placed inside a microwave cavity with a geometric size of $43.0\times 21.0\times9.6$~mm$^{3}$,  is coherently coupled to the electromagnetic mode in the cavity via the magnetic dipole interaction \cite{Tang2014}.  We assume that only the Kittel mode of the magnet is excited by the magnetic component of the microwave field.  The cavity can be excited by an external microwave source,  either a discrete pulse \cite{Tang2014,Hu2019PRB,Weides2020} or a continuous wave \cite{Huebl2013,Nakamura2014,Tang2014,Tobar2014,Hu2015-1}.  The Hamiltonian of this system is \cite{Tang2014,Li2018}
\begin{eqnarray}
H&=&\hbar \omega_{c}c^{\dag}c+\hbar\omega_{m}m^{\dag}m+\hbar g(c^{\dag}m+m^{\dag}c)\nonumber\\
&&+i\hbar\Omega(e^{-i\omega_{0}t}c^{\dag}-e^{i\omega_{0}t}c).\label{Ham}
\end{eqnarray}
Here,  $\hbar$ is the reduced Planck constant; $c^{\dag}$($c$) and $m^{\dag}$($m$) are the creation (annihilation) operators of the cavity photons and magnons with eigenfrequencies $\omega_{c}$ and $\omega_{m}$, respectively; $g$ is the coupling rate between the cavity photons and magnons,  where the rotating-wave approximation (RWA) has been employed;  and $\Omega$ and $\omega_{0}$ are the strength and frequency of the continuous driving microwave.  The last term in Eq.~(\ref{Ham}) will not be included for the pulse excitation scenario. 

In order to incorporate the dissipation effect microscopically,  we assume that the cavity photons and magnons are independently coupled to a corresponding thermal bath\cite{Leggett1983}.  The dynamics of this system will then be governed by the quantum master equation for the reduced density matrix $\rho$ (Appendix A)
\begin{eqnarray}
\frac{d\rho}{dt}=\frac{1}{i\hbar}[H,\rho]+\mathcal{L}\{\rho\}.\label{master}
\end{eqnarray}
Here,  $\mathcal{L}\{\rho\}$ is the Lindblad operator,  where $\mathcal{L}\{\rho\}=\sum\limits_{o=c,m}[\kappa_{o}(1+\overline{n}_{o})(2o\rho o^{\dag}-\{o^{\dag}o,\rho\})+\kappa_{o}\overline{n}_{o}(2o^{\dag}\rho o-\{oo^{\dag},\rho\})]$; $\overline{n}_{c}$ ( $\overline{n}_{m}$) is the average number of thermal cavity photons (magnons) with frequency $\omega_{c}$ ($\omega_{m}$) for the thermal bath with temperature $T$; and $\kappa_{c}$ and $\kappa_{m}$  are the damping rates for the cavity photons and magnons.   Once Eq.~(\ref{master}) is solved,  the correlation functions $\langle c^{\dag p}c^{q}m^{\dag r}m^{s}\rangle$ to arbitrary order $(p,q,r,s)$ of the operators $c^{\dag},c,m^{\dag},m$ can be obtained for the cavity magnon-polariton system.  Here,  $\langle\mathcal{O}\rangle$,  taking the form $\text{Tr}[\rho\mathcal{O}]$,  is the quantum statistical average of a generic operator $\mathcal{O}$ over the density matrix $\rho$.

In terms of the coherent states $|\alpha\rangle$ and $|\beta\rangle$ for the cavity photons and magnons,  respectively,  the density matrix $\rho$ can be expressed as $\rho=\int d^{2}\alpha d^{2}\beta\mathcal{P}(\alpha,\beta)|\alpha,\beta\rangle\langle\alpha,\beta|$.  The quasiprobability distribution function $\mathcal{P}(\alpha,\beta)$ will then satisfy the Fokker-Planck equation (Appendix B)
\begin{eqnarray}
\frac{\partial\mathcal{P}}{\partial t}&=&i\omega_{c}\frac{\partial}{\partial\alpha}(\alpha\mathcal{P})+i\omega_{m}\frac{\partial}{\partial\beta}(\beta\mathcal{P})+ig(\beta\frac{\partial}{\partial\alpha}+\alpha\frac{\partial}{\partial\beta})\nonumber\\
&-&\Omega e^{-i\omega_{0}t}\frac{\partial}{\partial\alpha}\mathcal{P}+\kappa_{c}\frac{\partial}{\partial\alpha}(\alpha\mathcal{P})+\kappa_{m}\frac{\partial}{\partial\beta}(\beta\mathcal{P})\nonumber\\
&+&\kappa_{c}\overline{n}_{c}\frac{\partial^{2}}{\partial\alpha\partial\alpha^{*}}\mathcal{P}+\kappa_{m}\overline{n}_{m}\frac{\partial^{2}}{\partial\beta\partial\beta^{*}}\mathcal{P}
+h.c. \label{FPE}
\end{eqnarray}
Then the expectation value $\mathcal{O}$ can be further expressed as
\begin{eqnarray}
\langle\mathcal{O}\rangle=\int d^{2}\alpha d^{2}\beta\mathcal{P}(\alpha,\beta,t)\langle\alpha,\beta|\mathcal{O}|\alpha,\beta\rangle.\label{avgO}
\end{eqnarray}

In addition to solving the Fokker-Planck equation (\ref{FPE}) directly,   the quasiprobability distribution function $\mathcal{P}(\alpha,\beta,t)$ can also be obtained by simulating the stochastic differential equations for $\alpha$ and $\beta$\cite{QuanOpt2}
\begin{eqnarray}
d\alpha&=&(-i\omega_{c}\alpha-ig\beta+\Omega e^{-i\omega_{0}t}-\kappa_{c}\alpha)dt\nonumber\\&&+\sqrt{\kappa_{c}\overline{n}_{c}}(dW_{1}+idW_{2}),\label{dalpha}\\
d\beta&=&(-i\omega_{m}\beta-ig\alpha-\kappa_{m}\beta)dt\nonumber\\&&+\sqrt{\kappa_{m}\overline{n}_{m}}(dW_{3}+idW_{4}),\label{dbeta}
\end{eqnarray}
Here, $dW_{i}(i=1,2,3,4)$ are independent Wiener processes, whose increasements satisfy the Gaussian distribution with expectation value $0$ and variance $dt$.  The statistical assembles of quantum trajectories of $\alpha$ and $\beta$ generated by Eq.~(\ref{dalpha}) and (\ref{dbeta}) will give the quasiprobability distribution function $\mathcal{P}(\alpha,\beta,t)$.  This so-called ``quantum trajectory method" can be more efficient from the computational aspect,  although these two methods are mathematically equivalent\cite{QuanOpt2}.  

Based on the Fokker-Planck equation (\ref{FPE}),  it is also able to get the dynamical equations for the correlation functions of cavity-photon and magnon operators.  For example,   the time derivative of $\langle c\rangle$ can be expressed as $\frac{\partial}{\partial t}\langle c\rangle=\int d^{2}\alpha d^{2}\beta \frac{\partial\mathcal{P}}{\partial t}\alpha$,  according to Eq.~(\ref{avgO}).  By substituting the expression of $\frac{\partial\mathcal{P}}{\partial t}$ given by Eq.~(\ref{FPE}) and performing the integrations over $\alpha$ and $\beta$,  one will get
\begin{eqnarray}
\frac{\partial}{\partial t}\langle c\rangle=-i\omega_{c}\langle c\rangle-ig\langle m\rangle-\kappa_{c}\langle c\rangle +\Omega e^{-i\omega_{0}t}.\label{alpha}
\end{eqnarray}   
Similary,  the equation for $\langle m\rangle$ will be
\begin{eqnarray}
\frac{\partial}{\partial t}\langle m\rangle=-i\omega_{m}\langle m\rangle-ig\langle c\rangle-\kappa_{m}\langle m\rangle.\label{beta}
\end{eqnarray}   
In fact,  Eq.~(\ref{alpha}) and (\ref{beta}) describe the dynamics of the coherent components in cavity photons and magnons respectively.  

The equations for arbitrary-order correlation functions $\langle c^{\dag p}c^{q}m^{\dag r}m^{s}\rangle$  can also be derived in the same way.  Using the fact $\frac{\partial\langle\mathcal{O}\rangle}{\partial t}=\int d^{2}\alpha d^{2}\beta\frac{\partial\mathcal{P}}{\partial t}(\alpha,\beta,t)\langle\alpha,\beta|\mathcal{O}|\alpha,\beta\rangle$,  we have got 
\begin{widetext}
\begin{eqnarray}
\frac{\partial}{\partial t}\langle c^{\dag p}c^{q}m^{\dag r}m^{s}\rangle&=&
[i(p-q)\omega_{c}-\kappa_{c}(p+q)+i(r-s)\omega_{m}-\kappa_{m}(r+s)]\langle c^{\dag p}c^{q}m^{\dag r}m^{s}\rangle\nonumber\\
&+&ig(p\langle c^{\dag p-1}c^{q}m^{\dag r+1}m^{s}\rangle-q\langle c^{\dag p}c^{q-1}m^{\dag r}m^{s+1}\rangle+r\langle c^{\dag p+1}c^{q}m^{\dag r-1}m^{s}\rangle-s\langle c^{\dag p}c^{q+1}m^{\dag r}m^{s-1}\rangle)\nonumber\\
&+&p\Omega e^{i\omega_{0}t}\langle c^{\dag p-1}c^{q}m^{\dag r}m^{s}\rangle+q\Omega e^{-i\omega_{0}t}\langle c^{\dag p}c^{q-1}m^{\dag r}m^{s}\rangle\nonumber\\
&+&2pq\kappa_{c}\overline{n}_{c}\langle c^{\dag p-1}c^{q-1}m^{\dag r}m^{s}\rangle+2rs\kappa_{m}\overline{n}_{m}\langle c^{\dag p}c^{q}m^{\dag r-1}m^{s-1}\rangle.
\label{recurr}
\end{eqnarray}
\end{widetext}
Notice that $p,q,r,s$ are non-negative integers,  and the terms with negative exponents on the right-hand side of Eq.~(\ref{recurr}) should vanish.  Therefore,  the higher-order correlation functions will be dependent on the lower-order correlation functions hierarchically.  

The Fokker-Planck equation (\ref{FPE}),  stochastic differential equations (\ref{dalpha})(\ref{dbeta}),  and the hierarchical equations (\ref{recurr}) are the central results to describe the driven-dissipative quantum dynamics in cavity magnon-polariton systems.  In the following,   this theoretical approach will be exploited to investigate the average number and second-order quantum coherence for cavity photons and magnons in two experimental scenarios.

\section{Results and Discussions}
\subsection{Average Number of Cavity Photons and Magnons} Existing experiments on the cavity magnon-polariton system have focused on the power spectrum of the microwave field,  which is proportional to the average number of microwave photons $\langle c^{\dag}c\rangle$ in the cavity.   By setting suitable integers $(p,q,r,s)$ for Eq.~(\ref{recurr}),  a group of coupled equations for $\langle c^{\dag}c\rangle$,  $\langle m^{\dag}m\rangle$,  $\langle cm^{\dag}\rangle$,  $\langle c^{\dag}m\rangle$,  $\langle c^{\dag}\rangle$,  $\langle c\rangle$,  $\langle m^{\dag}\rangle$,  and $\langle m\rangle$ can be obtained as
\begin{eqnarray}
\frac{\partial}{\partial t}\langle c^{\dag}c\rangle&=&-ig\langle c^{\dag}m\rangle +ig\langle cm^{\dag}\rangle+\Omega e^{-i\omega_{0}t}\langle c^{\dag}\rangle+\Omega e^{i\omega_{0}t}\langle c\rangle\nonumber\\&&-2\kappa_{c}\langle c^{\dag}c\rangle +2\kappa_{c}\overline{n}_{c},\label{alphaalpha}\\
\frac{\partial}{\partial t}\langle m^{\dag}m\rangle&=&ig\langle c^{\dag}m\rangle-ig\langle cm^{\dag}\rangle -2\kappa_{m}\langle m^{\dag}m\rangle+2\kappa_{m}\overline{n}_{m}, \label{betabeta}\\
\frac{\partial}{\partial t}\langle c^{\dag}m\rangle&=&i\omega_{c}\langle c^{\dag}m\rangle -i\omega_{m}\langle c^{\dag}m\rangle -ig \langle c^{\dag}c\rangle+ig\langle m^{\dag}m\rangle\nonumber\\
&&+\Omega e^{i\omega_{0}t}\langle m\rangle-\kappa_{c}\langle c^{\dag}m\rangle-\kappa_{m}\langle c^{\dag}m\rangle, \label{alphabeta}\\
\frac{\partial}{\partial t}\langle cm^{\dag}\rangle&=&-i\omega_{c}\langle cm^{\dag}\rangle+i\omega_{m}\langle cm^{\dag}\rangle
+ig\langle c^{\dag}c\rangle-ig\langle m^{\dag}m\rangle\nonumber\\&&+\Omega e^{-i\omega_{0}t}\langle m^{\dag}\rangle-\kappa_{c}\langle cm^{\dag}\rangle-\kappa_{m}\langle cm^{\dag}\rangle.\label{betaalpha}
\end{eqnarray}
One can see that the average number of thermal cavity photons and magnons will be involved here.  

If the cavity is continuously driven by the external microwave field,  the dynamics of the system will become stationary after a long time.  In this experimental scenario,  the solution of Eq.~(\ref{alpha}) and (\ref{beta}) can be written as $\langle c\rangle(t)=\alpha_{0}e^{-i\omega_{0}t}$ and $\langle m\rangle(t)=\beta_{0}e^{-i\omega_{0}t}$,  with the amplitudes
\begin{eqnarray}
\alpha_{0}&=&-\frac{i\Omega(\omega_{0}-\omega_{m}+i\kappa_{m})}{(\omega_{+}-\omega_{0})(\omega_{-}-\omega_{0})},\label{alpha0}\\
\beta_{0}&=& -\frac{i\Omega g}{(\omega_{+}-\omega_{0})(\omega_{-}-\omega_{0})}.\label{beta0}
\end{eqnarray}
Here,  $\omega_{\pm}$ are the eigen frequencies of the two branches of cavity magnon-polariton modes,  where $\omega_{\pm}=\frac{\omega_{c}+\omega_{m}}{2}-i\frac{\kappa_{c}+\kappa_{m}}{2}\pm\sqrt{(\frac{\omega_{c}-\omega_{m}}{2}-i\frac{\kappa_{c}-\kappa_{m}}{2})^{2}+g^{2}}$.  Furthermore,  the average number of cavity photons and magnons will become constant in this case.   Specially,  the solutions of Eq.~(\ref{alphaalpha})(\ref{betabeta}) will give
\begin{eqnarray}
\langle c^{\dag}c\rangle&=&|\alpha_{0}|^{2}+(1-\gamma_{m})\overline{n}_{c}+\gamma_{m}\overline{n}_{m}, \label{nphoton}\\
\langle m^{\dag}m\rangle&=&|\beta_{0}|^{2}+(1-\gamma_{c})\overline{n}_{m}+\gamma_{c}\overline{n}_{c},\label{nmagnon}   
\end{eqnarray}   
where $\gamma_{m}$ and $\gamma_{c}$ are 
\begin{eqnarray}
\gamma_{m}&=&\frac{g^{2}\kappa_{m}(\kappa_{c}+\kappa_{m})}{g^{2}(\kappa_{c}+\kappa_{m})^{2}+\kappa_{c}\kappa_{m}(\kappa_{c}+\kappa_{m})^{2}+\kappa_{c}\kappa_{m}(\omega_{m}-\omega_{c})^{2}},\nonumber\\
\gamma_{c}&=&\frac{g^{2}\kappa_{c}(\kappa_{c}+\kappa_{m})}{g^{2}(\kappa_{c}+\kappa_{m})^{2}+\kappa_{c}\kappa_{m}(\kappa_{c}+\kappa_{m})^{2}+\kappa_{c}\kappa_{m}(\omega_{m}-\omega_{c})^{2}}.\nonumber
\end{eqnarray}
One can find that both the drive source and thermal bath will affect the average number of cavity photons and magnons.   Fig.~\ref{Fig1}(b) further shows $\langle c^{\dag}c\rangle$ as a function of the bias magnetic field $B$ and the driving frequency $\omega_{0}$.   The anticrossing of two branches of the cavity photons clearly indicates the formation of two magnon-polariton modes,  which has been widely observed in previous experiments \cite{Review}.

Besides the level repulsion observed in the continuous drive scenario,  Rabi-like oscillation behavior is also observed in the microwave power spectrum in the pulse excitation scenario \cite{Tang2014,Hu2019PRB,Weides2020}.  Fig.~(\ref{Fig1})(c) shows the transient dynamics of $\langle c^{\dag}c\rangle$ in the cavity after a short pulsive excitation,  which have been obtained by solving the Fokker-Planck equation (\ref{FPE}) based on the quantum trajectory method.  The system parameters are the same as in Fig.~\ref{Fig1}(b) without the driving term,  and the initial excitation is taken into account by injecting $10^{8}$ coherent microwave photons.  For a rectangular pulse with frequency $\omega_{0}/2\pi=7.875$~GHz and duration $1$~ns,  the corresponding microwave power is estimated to be $-32.8$~dBm.  

The Rabi-like oscillation of $\langle c^{\dag}c\rangle$ in Fig.~\ref{Fig1}(c) can be simply understood from Eq.~(\ref{alpha}) and (\ref{beta}) by setting $\Omega=0$.  With the initial conditions $\langle c\rangle(0)=\langle c\rangle_{0},\langle m\rangle(0)=0$,  the solution of $\langle c\rangle(t) $ and $\langle m\rangle(t) $ will be the linear combination of the two cavity magnon-polariton modes,  namely,
\begin{eqnarray}
\left(\begin{array}{c}\langle c\rangle(t) \\ \langle m\rangle(t)\end{array}\right)=\sum_{i=+,-}\gamma_{i}\left(\begin{array}{c}\alpha_{i} \\\beta_{i}\end{array}\right)e^{-i\omega_{i} t}
.\label{ctmt}
\end{eqnarray}
Here,  $\left(\begin{array}{c}
\alpha_{\pm} \\ \beta_{\pm}
\end{array}\right)=\frac{1}{\sqrt{(\omega_{\pm}-\omega_{c})^{2}+\kappa_{c}^{2}+g^{2}}}\left(\begin{array}{c}
g \\ \omega_{\pm}-\omega_{c}+i\kappa_{c}
\end{array}\right)$ are the two normalized modes of cavity magnon-polaritons,  and the coefficients $\gamma_{\pm}$  are determined to be $
\gamma_{\pm}=\frac{\pm\beta_{\mp}\langle c\rangle_{0}}{\alpha_{+}\beta_{-}-\alpha_{-}\beta_{+}}$.   In the strong coupling case $\kappa_{c},\kappa_{m}\ll g$,  the contribution of $\langle c\rangle(t)$ and $\langle m\rangle(t)$ to the average number of cavity photons and magnons will approximately be
\begin{eqnarray}
|\langle c\rangle(t)|^{2}&=&|\langle c\rangle_{0}|^{2}(\mathcal{A}+\mathcal{B}+2\mathcal{C}\cos(\Delta\omega t))e^{-2\kappa t},\label{c2t}\\
|\langle m\rangle(t)|^{2}&=&|\langle c\rangle_{0}|^{2}\mathcal{C}(2-2\cos(\Delta\omega t) )e^{-2\kappa t}.\label{m2t}
\end{eqnarray}
Here,  $\Delta\omega=\sqrt{(\omega_{c}-\omega_{m})+4g^{2}}$ is the frequency difference between the two cavity magnon-polariton modes; and $\kappa= \frac{\kappa_{c}+\kappa_{m}}{2}$ is the avarage damping rate of the whole system.  The coefficients $\mathcal{A},\mathcal{B},\mathcal{C}$ are expressed as
$\mathcal{A}=\frac{|\alpha_{+}\beta_{-}|^{2}}{|\alpha_{+}\beta_{-}-\alpha_{-}\beta_{+}|^{2}},  \mathcal{B}=\frac{|\alpha_{-}\beta_{+}|^{2}}{|\alpha_{+}\beta_{-}-\alpha_{-}\beta_{+}|^{2}},
\mathcal{C}=\frac{|\alpha_{+}\alpha_{-}|^{2}}{|\alpha_{+}\beta_{-}-\alpha_{-}\beta_{+}|^{2}}=\frac{|\beta_{+}\beta_{-}|^{2}}{|\alpha_{+}\beta_{-}-\alpha_{-}\beta_{+}|^{2}}.
$

Eq.~(\ref{c2t}) and (\ref{m2t}) suggest that the Rabi-like oscillation of cavity photon numbers is caused by the interference effect between the two dissipative magnon-polariton modes.   The oscillation frequency depends on the frequency difference between these two modes,  which is at a minimum for the zero-detuned bias magnetic field $B=281.25$~mT.   Due to the dissipation effect,  the injected microwave photons will gradually decay and the whole system will reach thermal equilibrium.  The oscillatory decay of $\langle c^{\dag}c\rangle$ and $\langle m^{\dag}m\rangle$ at zero detuning are further plotted in Fig.~\ref{Fig1}(d),  which shows the interconversion between the average number of cavity photons and magnons.  The sharp dips suggest that nearly all the cavity photons will be converted into magnons.  It is noted that the proportion of cavity photons participating in the oscillation reaches a maximum when $\omega_{m}=\omega_{c}$,  and it will reduce with a strongly-detuned bias magnetic field.   This is characterized by the ratio $\frac{2\mathcal{C}}{\mathcal{A}+\mathcal{B}}$,  which will be $100\%$ at the zero-detuned bias magnetic field and will become smaller at larger-detuned bias magnetic field.  

\begin{figure}[!ht]
  \centering
  \includegraphics[width=0.50\textwidth,clip]{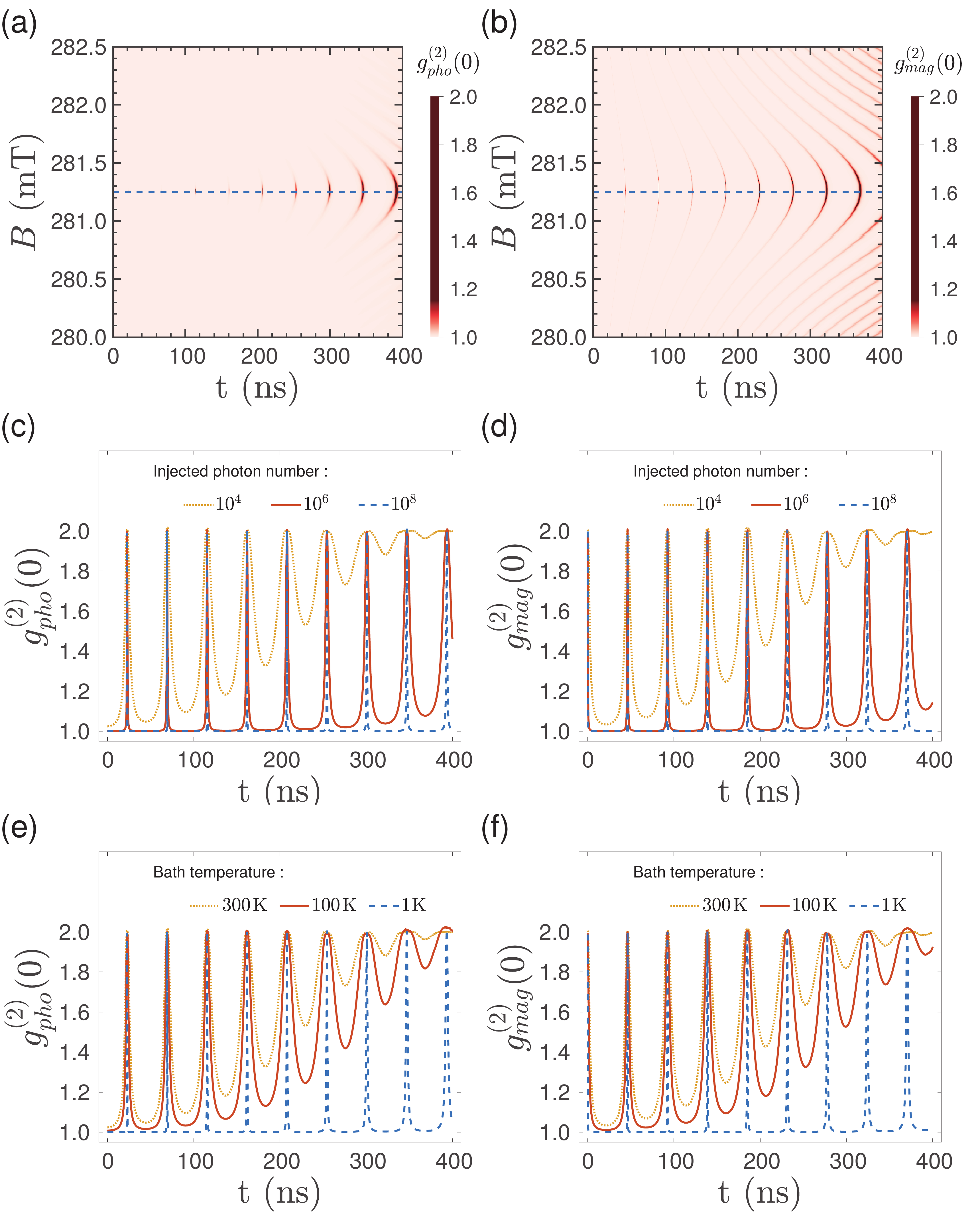}
  \caption{(Color online) The transient dynamics of the second-order quantum coherence after an initial pulse excitation.  (a)(b) $g_{pho}^{(2)}(0)$ and $g_{mag}^{(2)}(0)$ as functions of the bias magnetic field $B$,  with $10^{8}$ injected microwave photons and bath temperature $T=300$~K.  (c)(d) The time-evolution of $g_{pho}^{(2)}(0)$ and $g_{mag}^{(2)}(0)$ for different numbers of injected microwave photons,  with a zero-detuned bias magnetic field $B=281.25$~mT and bath temperature $T=300$~K.  (e)(f) The time-evolution of $g_{pho}^{(2)}(0)$ and $g_{mag}^{(2)}(0)$ for different bath temperatures,  with the zero-detuned bias magnetic field $B=281.25$~mT and $10^{4}$ microwave photons injected.  The other simulation parameters are the same as in Fig.~\ref{Fig1}(c) and (d). } 
 \label{Fig2}
\end{figure}

\subsection{Second-order Quantum Coherence} The cavity photons (magnons) will be either ``coherent" or ``incoherent" depending on whether their phases are locked to give non-zero $\langle c\rangle$ ($\langle m\rangle$) or not \cite{Review}.  Although the formation of cavity magnon-polaritons has been confirmed by observing level repulsion and Rabi-like oscillation in the microwave power spectrum,  no information about the coherent and incoherent components of the cavity photons or magnons in these dynamical processes can be extracted by merely measuring the first-order correlation function.  Hence,  we further investigate the second-order quantum coherence of the cavity photons and magnons,  which are characterized by the functions $g_{pho}^{(2)}(0)=\langle c^{\dag}c^{\dag}cc\rangle/\langle c^{\dag}c\rangle^{2}$, $g_{mag}^{(2)}(0)=\langle m^{\dag}m^{\dag}mm\rangle/\langle m^{\dag}m\rangle^{2}$.  The $g^{(2)}$ function has been extensively used to characterize the intensity correlation for a quantum optical field \cite{QuanOpt1,QuanOpt2,QuanOpt3}.  In particular,  a single-mode thermal field will have $g^{(2)}(0)=2$ and an optical field in the coherent state will have $g^{(2)}(0)=1$,  even though the two fields may have the same average number of photons.  Furthermore,  $g^{(2)}(0)$ can be less than $1$ for non-classical light \cite{QuanOpt1,QuanOpt2,QuanOpt3}.  Measuring the second-order quantum coherence could certainly provide indispensable knowledge about the quantum dynamics in the cavity magnon-polariton system.  

Fig.~\ref{Fig2}(a) and (b) show the transient dynamics of $g_{pho}^{(2)}(0)$ and $g_{mag}^{(2)}(0)$ after a pulse excitation as a function of the bias magnetic field $B$,  which are calculated from the quantum trajectory method with the same parameters in Fig.~1(c).  Unlike the oscillations of $\langle c^{\dag}c\rangle$ and $\langle m^{\dag}m\rangle$,  which reflect the exchange of energy between cavity photons and magnons,  the oscillations of $g_{pho}^{(2)}(0)$ and $g_{mag}^{(2)}(0)$ indicate the periodic modulation of these two bosonic fields between the coherent state and thermal state.  After the cavity photons are coherently excited by the initial pulse,  the coherent component between the cavity photons and magnons will be interconverted due to their strong coupling,  as suggested by Eq.~(\ref{c2t}) and (\ref{m2t}).  The oscillation in Fig.~\ref{Fig2}(a) and (b) is not obviously seen in the beginning,  where the coherent component is dominant over the incoherent component,  until it is reduced by the dissipation effect of the thermal bath.  Fig.~\ref{Fig2}(a) further shows that the coherent component of the cavity photons will be dominant for a longer time under a more-detuned bias magnetic field,  since only a small proportion of it will be involved in the interconversion with the magnons.  Therefore,  significant oscillation of second-order quantum coherence will become obvious only if the coherent component is comparable to the thermal counterpart in the two bosonic fields.   

\begin{figure}[!ht]
  \centering
  \includegraphics[width=0.50\textwidth,clip]{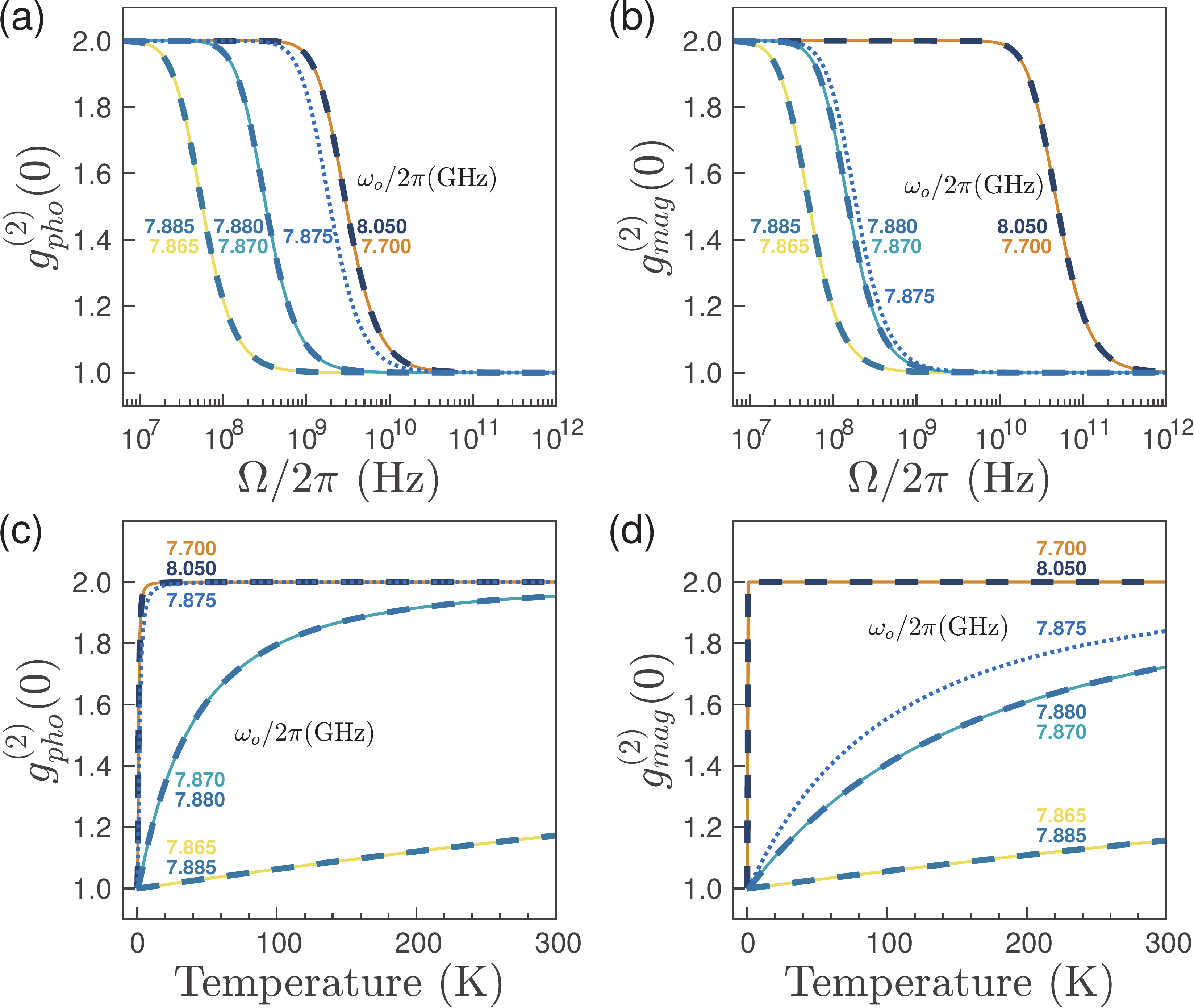}
  \caption{(Color online)  The second-order quantum coherence in the continuous drive scenario.  (a)(b) $g_{pho}^{(2)}(0)$ and $g_{mag}^{(2)}(0)$ as functions of the driving strength for different driving frequencies,  with a  zero-detuned bias magnetic field $B=281.25$~mT and bath temperature $T=300$~K.  (c)(d)  $g_{pho}^{(2)}(0)$ and $g_{mag}^{(2)}(0)$ as functions of the bath temperature for different driving frequencies,  with a zero-detuned bias magnetic field $B=281.25$~mT and driving strength $\Omega/2\pi=10^{8}$~Hz.  The other simulation parameters are the same as those in Fig.~\ref{Fig1}(b).  } 
 \label{Fig3}
\end{figure}

\begin{figure*}[!ht]
  \centering
  \includegraphics[width=1.0\textwidth,clip]{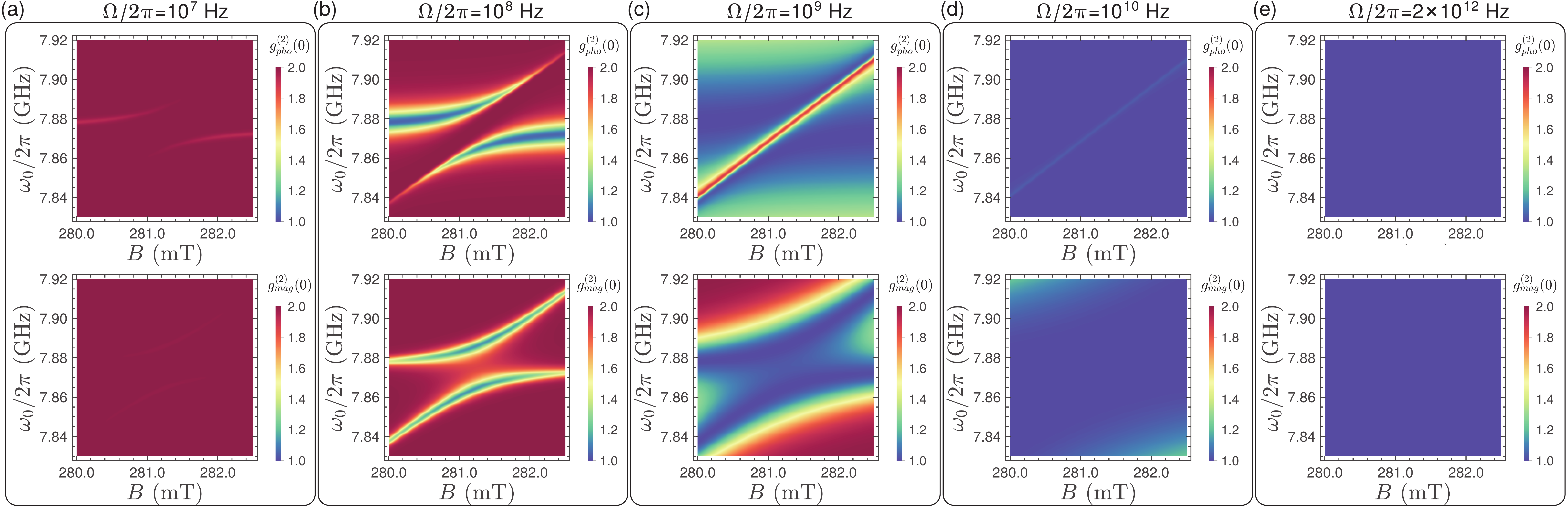}
  \caption{(Color online) Second-order coherence function $g_{pho}^{(2)}(0)$ and $g_{mag}^{(2)}(0)$ as functions of bias magnetic field $B$ and driving frequency $\omega_{0}$ in the continuous drive scenario,  with five different driving strengths.  (a) $\Omega/2\pi=10^{7}$~Hz; (b) $\Omega/2\pi= 10^{8}$~Hz;  (c) $\Omega/2\pi= 10^{9}$~Hz; (d) $\Omega/2\pi= 10^{10}$~Hz; (e) $\Omega/2\pi= 2\times 10^{12}$~Hz.  Here,  the bath temperature is fixed as $T=300$~K.  The other simulation parameters are the same as those in Fig.~\ref{Fig1}(b). } 
 \label{Fig4}
\end{figure*}

\begin{figure*}[!ht]
  \centering
  \includegraphics[width=1.0\textwidth,clip]{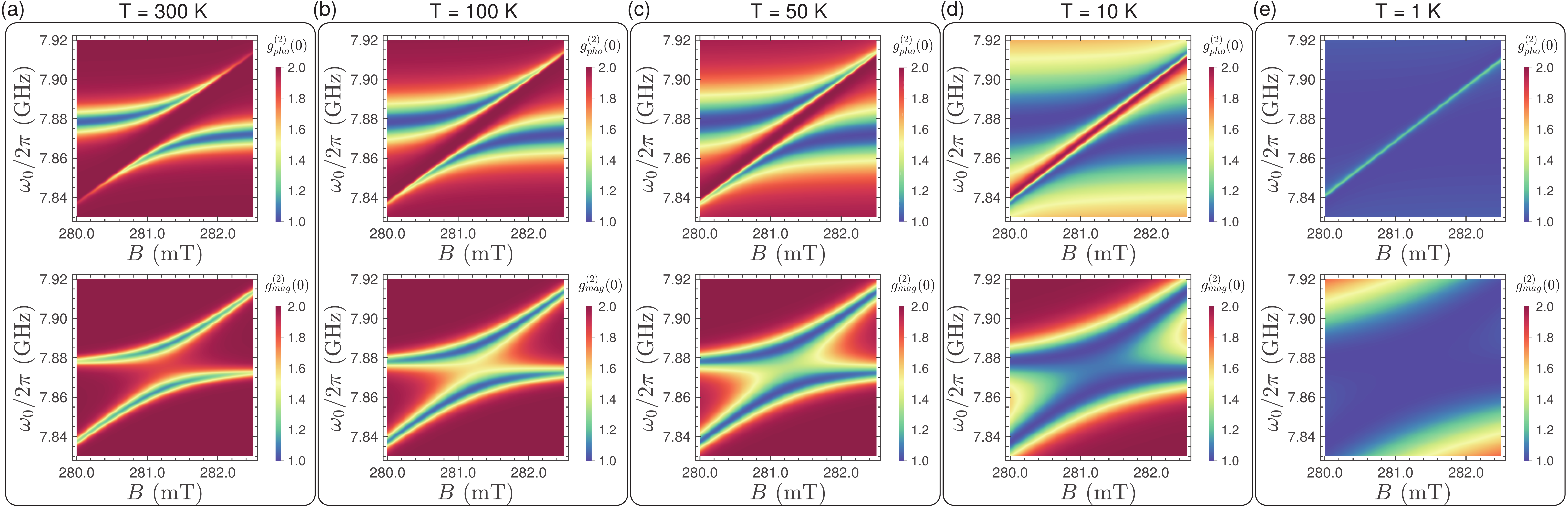}
  \caption{(Color online) Second-order coherence function $g_{pho}^{(2)}(0)$ and $g_{mag}^{(2)}(0)$ as functions of bias magnetic field $B$ and driving frequency $\omega_{0}$ in the continuous drive scenario,  with five different bath temperatures. (a) $T=300$~K; (b) $T= 100$~K;  (c) $T= 50$~K; (d) $T= 10$~K; (e) $T= 1$~K.  Here,  the driving strength is fixed as  $\Omega/2\pi= 10^{8}$~Hz. The other simulation parameters are the same as those in Fig.~\ref{Fig1}(b). } 
 \label{Fig5}
\end{figure*}

The competition between coherent and incoherent components in the transient dynamics of the cavity magnon-polariton system is further investigated for the zero-detuned bias magnetic field $B=281.25$~mT.  With fixed bath temperature $T=300$~K,  Fig.~\ref{Fig2}(c) and (d) show the periodic modulations of the cavity photons and magnons between the coherent state and thermal state after $10^{8}$, $10^{6}$, or $10^{4}$ cavity photons are coherently injected.  The corresponding microwave power will be $-32.8,-52.8,-72.8$~dBm for a rectangular pulse with frequency $\omega_{0}/2\pi=7.875$~GHz and duration $1$~ns.  Initially,  the magnons are still in the thermal state with $g_{mag}^{(2)}(0)=2$,  which causes a peak at $t=0$ in Fig.~2(d).   The more microwave photons are injected,  the longer the time during which the coherent component can suppress the incoherent component,  as incidated by the larger number of cycles with purely coherent states of the cavity photons and magnons.   On the other hand,  the incoherent component of the two bosonic fields can be tuned by the thermal bath.  With $10^{4}$ injected photons,  more cycles with purely coherent photon and magnon states can be recovered by decreasing the bath temperature from $300$~K to $1$~K(see Fig.~\ref{Fig2}(e) and (f)).   Furthermore,  the maxima of $g_{pho}^{(2)}(0)$ and $g_{mag}^{(2)}(0)$ have the same time offset as $\langle c^{\dag}c\rangle$ and $\langle m^{\dag}m\rangle$ in Fig.~\ref{Fig1}(d),  since only the coherent components of these two bosonic fields will participate in the interconversion process.  Our results thus demonstrate the dynamical control of second-order quantum coherence in the cavity magnon-polariton system by engineering the microwave pulse or thermal bath.  

The second-order quantum coherence in the continuous drive scenario can also be tuned by the drive source and thermal bath,  as shown in Fig.~3 by numerically solving Eq.~(\ref{recurr}) at the zero-detuned bias magnetic field.   Generally,  the two bosonic fields will transition from a thermal state to a coherent state when the driving strength is continuously increased and the coherent component becomes dominant (see Fig.~\ref{Fig3}(a) and (b)).  Moreover,  the critical driving strength for the transition is lowest if the driving frequency is in resonance with either the cavity magnon-polariton mode (7.865~or 7.885~GHz here).  For a given driving strength $\Omega/2\pi=10^{8}$~Hz,  Fig.~\ref{Fig3}(c) and (d) shows the transition from the coherent state to the thermal state if the bath temprature is increased and the incoherent component becomes dominant.  Once again,  the second-order quantum coherence is most robust against the thermal fluctuations when either cavity magnon-polariton mode is resonantly excited.  Furthermore,  the behaviors of the cavity photons and magnons are asymmetric if the driving frequency $\omega_{0}/2\pi=\omega_{c}/2\pi=\omega_{m}/2\pi=7.875$~GHz (see the dotted lines in Fig.~\ref{Fig3}(a)-(d)).  

The results in Fig.~\ref{Fig3} can be explained by the analytical expressions of the second-order quantum coherence,  where $g_{pho}^{(2)}=\frac{(|\alpha_{0}|^{2}+2\overline{n})^{2}-2\overline{n}^{2}}{(|\alpha_{0}|^{2}+\overline{n})^{2}},  g_{mag}^{(2)}=\frac{(|\beta_{0}|^{2}+2\overline{n})^{2}-2\overline{n}^{2}}{(|\beta_{0}|^{2}+\overline{n})^{2}}$.  Here,  $\alpha_{0}$ and $\beta_{0}$ are given by Eq.~(\ref{alpha0}) and (\ref{beta0}),  and one has $\overline{n}_{c}=\overline{n}_{m}\equiv\overline{n}$ at the zero-detuned bias magnetic field.  Therefore,  when $|\alpha_{0}|^{2}\gg \overline{n}$ and $|\beta_{0}|^{2}\gg \overline{n}$ for large drive strength or low bath temperature,  the second-order quantum coherence functions $g_{pho}^{(2)}(0)$ and $g_{mag}^{(2)}(0)$ will be nearly $1$; in contrast,  they will be nearly $2$ if $|\alpha_{0}|^{2}\ll \overline{n}$ and $|\beta_{0}|^{2}\ll \overline{n}$.  While for given driving strength and bath temperature,   $|\alpha_{0}|$ and $|\beta_{0}|$ will be largest when the $\omega_{0}=\omega_{+}$ or  $\omega_{0}=\omega_{-}$.  Besides, when $\omega_{0}=\omega_{m}$,  the ratio $|\alpha_{0}|/|\beta_{0}|=\kappa_{m}/g$,  which explains the asymmetric behavior between $g_{pho}^{(2)}(0)$ and $g_{mag}^{(2)}(0)$. 

We have also studied the dependence of second-order quantum coherence on the bias magnetic field and driving frequency with given driving strength or bath temperature,  as shown in Fig.~\ref{Fig4} and Fig.~\ref{Fig5} respectively.   When the driving strength is weak,  the incoherent components of cavity photons and magnons will be dominant,  and these two bosonic fields are nearly in the thermal state in the entire parameter space,  where $g_{pho}^{(2)}(0)\approx 2$ and $g_{mag}^{(2)}(0)\approx 2$ (see Fig.~\ref{Fig4}(a)).   With larger driving strength,  the coherent components of cavity photons and magnons will be enhanced,  especially when the driving frequency is resonant with the cavity magnon-polariton modes.  One can see that the level repulsion will also appear in the second-order quantum coherence (see Fig.~\ref{Fig4}(b)).  When the driving strength is further increased,  the coherent components will be dominant in more parameter space,  as shown in Fig.~\ref{Fig4}(c).   Finally,   the level repulsion will vanish again when the drive strengh is very strong,  since the cavity photons and magnons will be nearly in the coherent state in most parameter space,  where $g_{pho}^{(2)}(0)\approx 1$ and $g_{mag}^{(2)}(0)\approx 1$ (see Fig.~\ref{Fig4}(d)(f)).   On the other hand,  the feature of level repulsion will also be drastically modified when the bath temperature is continuously decreased (see Fig.~\ref{Fig5}).  As the incoherent components of cavity photons and magnons are suppressed at lower bath temperature,  these two bosonic fields will get closer to the coherent state.  Therefore,  the evolutions of level repulsion in the second-order quantum coherence have directly reflect the competition between the coherent and incoherent components in these two bosonic fields. 

\subsection{Experimental Proposal} 
\begin{figure}[!ht]
  \centering
  \includegraphics[width=0.5\textwidth,clip]{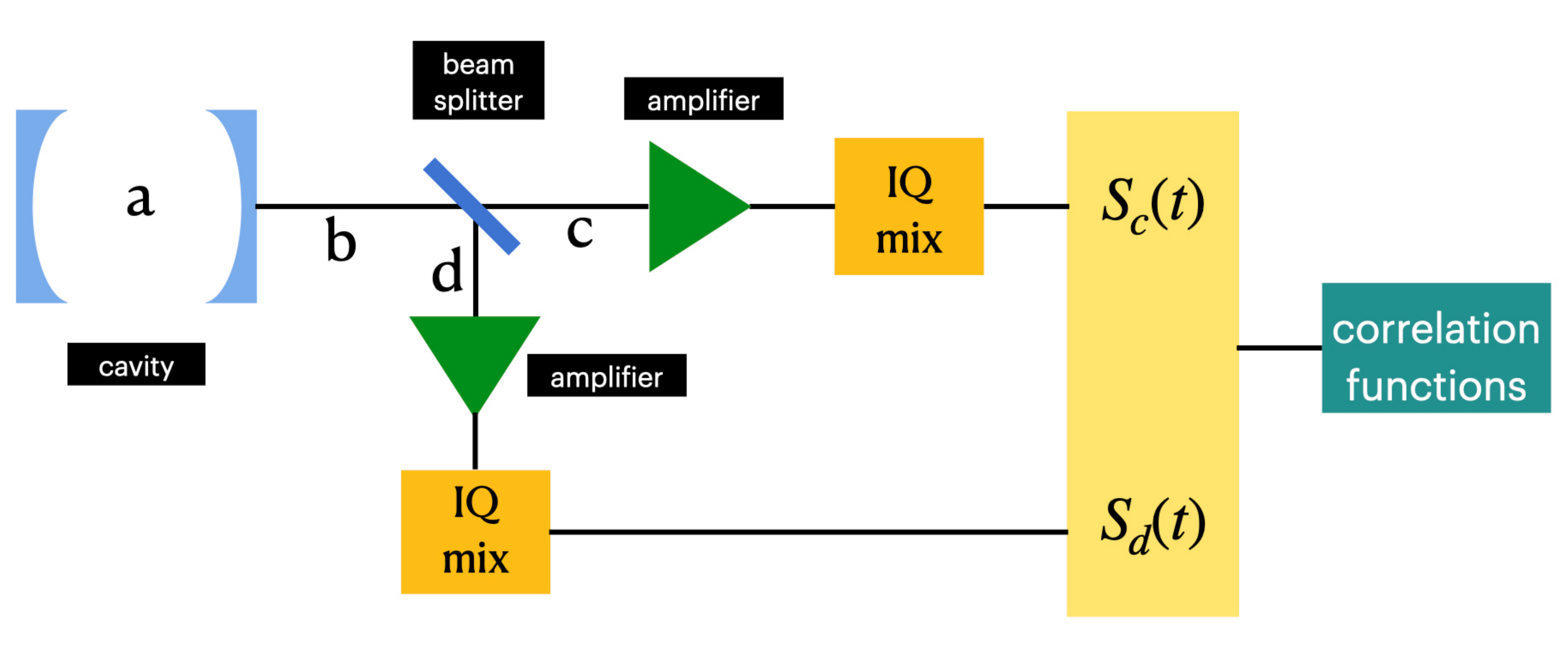}
  \caption{(Color online) A schematic diagram of the experimental setup to measure second-order quantum coherence of microwave field.  Here,  the microwave field $b$ emitted from the cavity is splitted into $c$ and $d$ by a $90^{\circ}$ hybrid coupler.  The two output modes $c$ and $d$ will be amplified first,  and then IQ mixers are used to perform the quadrature measurement on $c$ and $d$.   This will give the complex envelopes $S_{c}(t)$ and $S_{d}(t)$,  which can be used to extract the correlation functions of the microwave field\cite{daSilva2010}.} 
 \label{Fig6}
\end{figure}

Second-order quantum coherence is usually measured in the spirit of the Hanbury Brown-Twiss experimental setup \cite{HBT}.  For an optical field,  this can be performed with single-photon detectors.  However,  the detection of a single microwave photon is challenging,  since the energy of a microwave photon is about four or five orders of magnitude lower than that of an optical photon.  Instead,  experimental techniques with linear detectors \cite{daSilva2010} have been developed to measure the second-order coherence function of a microwave field in circuit QED systems \cite{Bozyigit2011,LangC2011,EichlerC2012,LangC2013, Peng2016,Gasparinetti2017,Rolland2019},  as schematically shown in Fig.~\ref{Fig6}.  Here,  a $90^{\circ}$ hybrid coupler is used as a beam splitter to separate the microwave field $b$ emitted from the cavity into two modes $c$ and $d$,  which will be amplified afterwards.  Then IQ mixers will be used to perform the quadrature measurement on $c$ and $d$ modes,  which gives the complex envelopes $S_{c}(t)$ and $S_{d}(t)$.   The correlation functions can be calculated from the measured $S_{c}(t)$ and $S_{d}(t)$\cite{daSilva2010}.  In the past,  this technique has been successfully applied to obtained the second-order correlation functions of microwave field in circuit QED systems\cite{Bozyigit2011,LangC2011,EichlerC2012,LangC2013, Peng2016,Gasparinetti2017,Rolland2019}.   We anticipate that the same experimental techniques can be exploited to investigate higher-order quantum correlation effects in cavity magnon-polariton systems. 

\section{Conclusion} In conclusion,  the driven-dissipative dynamics in a cavity magnon-polariton system has been theoretically studied in a full quantum level.  The Fokker-Planck equation,  stochastic differential equations,  and a group of hierarchical equations have been established to give the arbitrary-order correlation functions of cavity photons and magnons.  The theoretical approach has successfully reproduced the remarkable phenomena of level repulsion and Rabi-like oscillation observed in the microwave power spectrum.  Furthermore,  the second-order coherence functions of cavity photons and magnons have been thoroughly investigated for two typical experimental scenarios.  The results reveal the competition between the coherent and incoherent components in these two bosonic fields,  which can be systematically tuned by engineering the external drive source and the thermal bath.  Therefore,  measuring  second-order quantum coherence with currently available experimental techniques could provide another window to observe the rich dynamics in this hybrid quantum system.  It would also be interesting to search for nonclassical states for cavity photons and magnons by extending the current work to the dissipative coupling,  ultrastrong coupling,  or nonlinearly interacting cases,  which are of fundamental and practical importance for quantum cavity magnonics.  

\begin{acknowledgments}
G. G. Z.  and Y. W.  acknowledge support from NSFC Projects No.  61674083 and No.  12074195.  X.F.  Q.  acknowledges partial support from NSF PHY-1505189 and the Stevens Institute of Technology.
\end{acknowledgments}

\appendix
\section{Quantum Master Equation}
In this section,  we derive the quantum master equation for the reduced density matrix of the cavity magnon-polariton system.  The dissipation of the cavity photons and magnons will arise when they are coupled to the thermal bath.   We write the Hamiltonan for the whole system as
\begin{eqnarray}
\mathcal{H}=H+H_{B}+V,\label{HamW}
\end{eqnarray}
where 
\begin{eqnarray}
H&=&\hbar\omega_{c}c^{\dag}c+\hbar\omega_{m}m^{\dag}m+\hbar gc^{\dag}m+\hbar gm^{\dag}c\nonumber\\
&&+i\hbar\Omega(c^{\dag}e^{-i\omega_{0}t}-ce^{i\omega_{0}t}),\nonumber\\
H_{B}&=&\sum_{i}\hbar\omega_{i}a_{i}^{\dag}a_{i}+\sum_{j}\hbar\omega_{j}b_{j}^{\dag}b_{j},\nonumber\\
V&=&\sum_{i}g_{c,i}(c^{\dag}a_{i}+a_{i}^{\dag}c)+\sum_{j}g_{m,j}(m^{\dag}b_{j}+b_{j}^{\dag}m).\nonumber
\end{eqnarray} 
Here,  $H$ describes the cavity magnon-polariton system under consideration;  $H_{B}$ describes the thermal bath for cavity photons and magnons respectively,  which consists of an infinite set of harmonic oscillators with frequencies $\{\omega_{a,i}\}$ and $\{\omega_{b,j}\}$; and $V$ describes the coupling interaction between the system and the thermal bath.  The total density matrix $\rho_{T}$ for the whole system will satisfy the Liouville-von Neumann equation
\begin{eqnarray}
\frac{d\rho_{T}(t)}{dt}&=&\frac{1}{i\hbar}[\mathcal{H}(t),\rho_{T}(t)]\nonumber\\
&=&\frac{1}{i\hbar}[\mathcal{H}(t),\rho_{T}(0)]+(\frac{1}{i\hbar})^{2}\int_{0}^{t}[\mathcal{H}(t),[\mathcal{H}(\tau),\rho_{T}(\tau)]d\tau.\nonumber\\\label{rhot1}
\end{eqnarray}
In the interaction picture,   Eq.~(\ref{rhot1}) will become
\begin{eqnarray}
\frac{d\rho_{T}^{int}(t)}{dt}&=&\frac{1}{i\hbar}[\mathcal{V}(t),\rho_{T}^{int}(0)]\nonumber\\
&&+(\frac{1}{i\hbar})^{2}\int_{0}^{t}[\mathcal{V}(t),[\mathcal{V}(\tau),\rho_{T}^{int}(\tau)]d\tau.\nonumber\\\label{rhot2}
\end{eqnarray}
Here,  we have denoted $\rho_{T}^{int}(t)=U^{\dag}(t,0)\rho_{T}(t)U(t,0)$ and $\mathcal{V}(t)=U^{\dag}(t,0)V(t)U(t,0)$ using the unitary evolution operator $U(t,0)=U_{S}(t,0)U_{B}(t,0)$,  where $U_{S}(t,0)=e^{\frac{1}{i\hbar}\int_{0}^{t}H(\tau)d\tau},  U_{B}(t,0)= e^{\frac{1}{i\hbar}\int_{0}^{t}H_{B}(\tau)d\tau}$.   

The reduced density matrix $\rho^{int}$ for the cavity magnon-polariton system can be obtained by tracing over the degree of freedom of the thermal bath,  namely,  $\rho^{int}=\text{Tr}_{B}[\rho_{T}^{int}]$.   Therefore, the quantum master equation for $\rho^{int}$ can be obtained from Eq.~(\ref{rhot2}) as
\begin{eqnarray}
\frac{d\rho^{int}}{dt}&=&\frac{1}{i\hbar}\text{Tr}_{B}[\mathcal{V}(t),\rho_{T}^{int}(0)]\nonumber\\
&+&(\frac{1}{i\hbar})^{2}\text{Tr}_{B}\int_{0}^{t}[\mathcal{V}(t),[\mathcal{V}(\tau),\rho_{T}^{int}(\tau)]d\tau.\label{drhodt}
\end{eqnarray}

With the Born approximation $\rho_{T}(t)=\rho(t)\otimes\rho_{B}(0)$,   the first term in the $r.h.s.$ of Eq.~(\ref{drhodt}) will be
\begin{eqnarray}
&&\text{Tr}_{B}[\mathcal{V}(t),\rho_{T}^{int}(0)]\nonumber\\
&=&\text{Tr}_{B}(U_{S}^{\dag}U_{B}^{\dag}V(t)\rho_{T}(0)U_{S}U_{B}-U_{S}^{\dag}U_{B}^{\dag}\rho_{T}(0)V(t)U_{S}U_{B})\nonumber\\
&=&\sum_{i}g_{c,i}(U_{S}^{\dag}c^{\dag}\rho U_{S}\text{Tr}_{B}[a_{i}\rho_{B}]+U_{S}^{\dag}c\rho U_{S}\text{Tr}_{B}[a_{i}^{\dag}\rho_{B}]\nonumber\\
&&-U_{S}^{\dag}\rho c^{\dag} U_{S}\text{Tr}_{B}[\rho_{B}a_{i}]-U_{S}^{\dag}\rho c U_{S}\text{Tr}_{B}[\rho_{B}a_{i}^{\dag}])\nonumber\\
&+&\sum_{j}g_{m,j}(U_{S}^{\dag}m^{\dag}\rho U_{S}\text{Tr}_{B}[b_{j}\rho_{B}]+U_{S}^{\dag}m\rho U_{S}\text{Tr}_{B}[b_{j}^{\dag}\rho_{B}]\nonumber\\
&&-U_{S}^{\dag}\rho m^{\dag} U_{S}\text{Tr}_{B}[\rho_{B}b_{j}]-U_{S}^{\dag}\rho m U_{S}\text{Tr}_{B}[\rho_{B}b_{j}^{\dag}]).\label{Term1}
\end{eqnarray}
This term will vanish because one has $\text{Tr}_{B}[a_{i}\rho_{B}]=\text{Tr}_{B}[a_{i}^{\dag}\rho_{B}]=\text{Tr}_{B}[b_{j}\rho_{B}]=\text{Tr}_{B}[b_{j}^{\dag}\rho_{B}]=0$ for the thermal bath. 

For the parameter range we will study here,  the coupling term and drive term in $H$ will be much smaller than the terms of cavity photons and magnons.  Therefore,  the time-evolution operator can be further approximated as $U_{S}(t,0)\approx e^{-i\omega_{c}c^{\dag}ct}e^{-i\omega_{m}m^{\dag}mt}$.  Then the second term in the $r.h.s$ of Eq.~(\ref{drhodt}) will describe the damping of the cavity photons and magnons due to the thermal bath individually,  which will give \cite{QuanOpt1}
\begin{eqnarray}
&&\frac{d\rho^{int}}{dt}\nonumber\\&=&-\kappa_{c}\overline{n}_{c}(cc^{\dag}\rho^{int}-2c^{\dag}\rho^{int}c+\rho^{int}cc^{\dag})\nonumber\\
&&-\kappa_{c}(\overline{n}_{c}+1)(c^{\dag}c\rho^{int}-2c\rho^{int}c^{\dag}+\rho^{int}c^{\dag}c)\nonumber\\
&&-\kappa_{m}\overline{n}_{m}(mm^{\dag}\rho^{int}-2m^{\dag}\rho^{int}m+\rho^{int}mm^{\dag})\nonumber\\
&&-\kappa_{m}(\overline{n}_{m}+1)(m^{\dag}m\rho^{int}-2m\rho^{int}m^{\dag}+\rho^{int}m^{\dag}m).\nonumber\\\label{Term2}
\end{eqnarray}
Here,  $\kappa_{c}$ ($\kappa_{m}$) is the damping rate for cavity photons (magnons),  and $\overline{n}_{c}$($\overline{n}_{m}$) is the average number of the quanta at frequency $\omega_{c}$($\omega_{m}$) in the thermal bath\cite{QuanOpt1}.  Transforming the result above back to the Schr\"{o}dinger picture,  we get the Lindblad form of the quantum master equation (\ref{master}) in the body text.

\section{Fokker-Planck Equation}
In this section,  we describe how to get the Fokker-Planck equation from the quantum master equation.  In the coherent state representation for cavity photons and magnons $|\alpha,\beta\rangle$,  the reduced density matrix $\rho$ can be expressed in terms of the quasi-probability distribution function $\mathcal{P}(\alpha,\beta)$ as\cite{QuanOpt1}
\begin{eqnarray}
\rho=\int d^{2}\alpha d^{2}\beta\mathcal{P}(\alpha,\beta)|\alpha,\beta\rangle\langle\alpha,\beta|.\label{Pdist}
\end{eqnarray}  
Substituting the expression (\ref{Pdist}) into the left and right sides of Eq.~(\ref{master}), we will have
\begin{widetext}
\begin{eqnarray}
&&\int d^{2}\alpha d^{2}\beta \frac{\partial\mathcal{P}}{\partial t} |\alpha,\beta\rangle \langle \alpha,\beta| \nonumber\\
&=&\int d^{2}\alpha d^{2}\beta \mathcal{P}\{ 
-i\omega_{c} (c^{\dag}c|\alpha,\beta\rangle \langle \alpha,\beta| - |\alpha,\beta\rangle \langle \alpha,\beta|c^{\dag}c )-i\omega_{m} (m^{\dag}m|\alpha,\beta\rangle \langle \alpha,\beta| - |\alpha,\beta\rangle \langle \alpha,\beta|m^{\dag}m)\nonumber \\
&&-g \left(c^{\dag}m|\alpha,\beta\rangle \langle \alpha,\beta| - |\alpha,\beta\rangle \langle \alpha,\beta|c^{\dag}m\right)-g \left(m^{\dag}c|\alpha,\beta\rangle \langle \alpha,\beta| - |\alpha,\beta\rangle \langle \alpha,\beta|m^{\dag}c\right)\nonumber \\
&&+\Omega (c^{\dag}e^{-i\omega_{0}t}|\alpha,\beta\rangle \langle \alpha,\beta|-ce^{i\omega_{0}t}|\alpha,\beta\rangle \langle \alpha,\beta|-|\alpha,\beta\rangle \langle \alpha,\beta|c^{\dag}e^{-i\omega_{0}t}+|\alpha,\beta\rangle \langle \alpha,\beta|ce^{i\omega_{0}t})\nonumber \\
&&-\kappa_{c}(1+\overline{n}_{c})(c^{\dag}c|\alpha,\beta\rangle \langle \alpha,\beta|-2c|\alpha,\beta\rangle \langle \alpha,\beta| c^{\dag}+|\alpha,\beta\rangle \langle \alpha,\beta| c^{\dag}c)\nonumber\\
&&-\kappa_{c}\overline{n}_{c}(|\alpha,\beta\rangle \langle \alpha,\beta| cc^{\dag}-2c^{\dag}|\alpha,\beta\rangle \langle \alpha,\beta| c+cc^{\dag}|\alpha,\beta\rangle \langle \alpha,\beta|)\nonumber \\
&&-\kappa_{m}(1+\overline{n}_{m})(m^{\dag}m|\alpha,\beta\rangle \langle \alpha,\beta|-2m|\alpha,\beta\rangle \langle \alpha,\beta| m^{\dag}+|\alpha,\beta\rangle \langle \alpha,\beta| m^{\dag}m)\nonumber\\&&-\kappa_{m}\overline{n}_{m}(|\alpha,\beta\rangle \langle \alpha,\beta| mm^{\dag}-2m^{\dag}|\alpha,\beta\rangle \langle \alpha,\beta| m+mm^{\dag}|\alpha,\beta\rangle \langle \alpha,\beta|)
\}.\label{FK1}
\end{eqnarray}
\end{widetext}
Using the following rules for the operators $c, c^{\dag}, m, m^{\dag}$ acting on the coherent state $|\alpha,\beta\rangle$\cite{QuanOpt1}
\begin{widetext}
\begin{eqnarray}
c|\alpha,\beta\rangle\langle \alpha,\beta|&=&\alpha|\alpha,\beta\rangle\langle \alpha,\beta|, |\alpha,\quad
\beta\rangle\langle \alpha,\beta|c^{\dag}=|\alpha,\beta\rangle\langle \alpha,\beta|\alpha^{*},\nonumber\\
m|\alpha,\beta\rangle\langle \alpha,\beta|&=&\beta|\alpha,\beta\rangle\langle \alpha,\beta|,\quad
|\alpha,\beta\rangle\langle \alpha,\beta|m^{\dag}=|\alpha,\beta\rangle\langle \alpha,\beta|\beta^{*},\nonumber\\
c^{\dag}|\alpha,\beta\rangle\langle \alpha,\beta|&=&(\frac{\partial}{\partial\alpha}+\alpha^{*})|\alpha,\beta\rangle\langle \alpha,\beta|,\quad
|\alpha,\beta\rangle\langle \alpha,\beta|c=(\frac{\partial}{\partial\alpha^{*}}+\alpha)|\alpha,\beta\rangle\langle \alpha,\beta|,\nonumber\\
m^{\dag}|\alpha,\beta\rangle\langle \alpha,\beta|&=&(\frac{\partial}{\partial\beta}+\beta^{*})|\alpha,\beta\rangle\langle \alpha,\beta|, \
|\alpha,\beta\rangle\langle \alpha,\beta|m=(\frac{\partial}{\partial\beta^{*}}+\beta)|\alpha,\beta\rangle\langle \alpha,\beta|,\nonumber
\end{eqnarray}
\end{widetext}
we will get the Fokker-Planck equation (\ref{FPE}) for $\mathcal{P}$ in the body text.


\begin{thebibliography}{99}
\bibitem{Imamoglu2009} A.  Imamo\u{g}lu, Phys. Rev. Lett. \textbf{102}, 083602 (2009).
\bibitem{Flatte2010-1} O. O. Soykal and M. E. Flatt\'{e}, Phys. Rev. Lett. \textbf{104}, 077202 (2010).
\bibitem{Flatte2010-2} O. O. Soykal and M. E. Flatt\'{e}, Phys. Rev. B \textbf{82}, 104413 (2010).
\bibitem{Huebl2013} H. Huebl, C. W. Zollitsch, J. Lotze, F. Hocke, M. Greifenstein, A. Marx, R. Gross, and S. T. B. Goennenwein, Phys. Rev. Lett. \textbf{111}, 127003 (2013).
\bibitem{Nakamura2014} Y. Tabuchi, S. Ishino, T. Ishikawa, R. Yamazaki, K. Usami, and Y. Nakamura, Phys. Rev. Lett. \textbf{113}, 083603 (2014).
\bibitem{Tang2014} X. Zhang, C.-L. Zou, L. Jiang, and H. X. Tang, Phys. Rev. Lett. \textbf{113}, 156401 (2014).
\bibitem{Tobar2014} M. Goryachev, W. G. Farr, D. L. Creedon, Y. Fan, M. Kostylev, and M. E. Tobar, Phys. Rev. Appl. \textbf{2}, 054002(2014).
\bibitem{Hu2015-1} L. Bai, M. Harder, Y. P. Chen, X. Fan, J. Q. Xiao, and C.-M. Hu, Phys. Rev. Lett. \textbf{114}, 227201 (2015). 
\bibitem{Review} B. Z.  Rameshti,  S. V.  Kusminskiy,  J.A.  Haigh,  K.  Usami,  D. Lachance-Quirion,  Y.  Nakamura,  C. -M.  Hu,  H. X.  Tang,  G.E.W. Bauer,  Y.M. Blanter,   arXiv:2106.09312.
\bibitem{Haigh2015} J. A. Haigh, N. J. Lambert, A. C. Doherty, and A. J. Ferguson, Phys. Rev. B \textbf{91}, 104410 (2015).
\bibitem{Hu2015-2} B. M. Yao, Y. S. Gui, Y. Xiao, H. Guo, X. S. Chen, W. Lu, C. L. Chien, and C.-M. Hu, Phys. Rev. B \textbf{92}, 184407 (2015).
\bibitem{Hu2018-NC} B. Yao, Y.S. Gui, J.W. Rao, S. Kaur, X.S. Chen, W. Lu, Y. Xiao, H. Guo, K.-P. Marzlin, and C.-M. Hu, Nat. Commun. \textbf{8}, 1437 (2018).
\bibitem{APL2020} C. A. Pottsa and J. P. Davis, Appl. Phys. Lett. \textbf{116}, 263503 (2020).  
\bibitem{Hu2016-APL} S. Kaur, B. M. Yao, J. W. Rao, Y. S. Gui, and C.-M. Hu, Appl. Phys. Lett. \textbf{109}, 032404 (2016). 
\bibitem{Tobar2018} M. Goryachev, S. Watt, J. Bourhill, M. Kostylev, and M. E. Tobar, Phys. Rev. B \textbf{97}, 155129 (2018).
\bibitem{Boventer2018} I. Boventer, M. Pfirrmann, J. Krause, Y. Sch\"{o}n, Mathias Kl\"{a}ui, and M. Weides, Phys. Rev. B \textbf{97}, 184420 (2018).
\bibitem{XFZhang2020} J. Xu, C. Zhong, X. Han, D. Jin, L. Jiang, and X. Zhang, Phys. Rev. Lett. \textbf{125}, 237201 (2020).
\bibitem{You2018} Y.-P. Wang, G.-Q. Zhang, D. Zhang, T.-F. Li, C.-M. Hu, and J. Q. You, Phys. Rev. Lett. \textbf{120}, 057202 (2018).
\bibitem{Hu2018-PRB} P. Hyde, B. M. Yao, Y. S. Gui, G.-Q. Zhang, J. Q. You, and C.-M. Hu, Phys. Rev. B \textbf{98}, 174423 (2018).
\bibitem{Hu2018-PRL}  M. Harder, Y. Yang, B. M. Yao, C. H. Yu, J. W. Rao, Y. S. Gui, R. L. Stamps, and C.-M. Hu, Phys. Rev. Lett. \textbf{121}, 137203 (2018). 
\bibitem{Xia2018} V. L. Grigoryan, K. Shen, and K. Xia, Phys. Rev. B \textbf{98}, 024406 (2018).
\bibitem{Xiao2019} W. Yu, J. Wang ,  H. Y. Yuan,  and J. Xiao, Phys. Rev. Lett. \textbf{123}, 227201 (2019).
\bibitem{You2018-NC} D. Zhang, X.-Q. Luo, Y.-P. Wang, T.-F. Li, and J.Q. You, Nat. Commun. \textbf{8}, 1368 (2018).
\bibitem{Yan2019} Y. Cao and P. Yan, Phys. Rev. B \textbf{99}, 214415 (2019).
\bibitem{Tang2015} X. Zhang, C.-L. Zou, N. Zhu, F. Marquardt, L. Jiang,  and H. X. Tang, Nat. Commun. \textbf{6}, 8914 (2015).
\bibitem{Hu2019} J.W. Rao, S. Kaur, B.M. Yao, E.R.J. Edwards, Y.T. Zhao, X. Fan, D.  Xue, T.J. Silva, Y.S. Gui, and C.-M. Hu,  Nat. Commun.  \textbf{10}, 2934 (2019).
\bibitem{Bai2017} L. Bai, M. Harder, P. Hyde, Z. Zhang, C.-M. Hu, Y. P. Chen, and J. Q. Xiao,  Phys. Rev. Lett. \textbf{118}, 217201 (2017).
\bibitem{WuY2019} Z.-X.  Liu,  H.  Xiong,  and Y.  Wu,  Phys.  Rev.  B \textbf{100},  134421 (2019).  
\bibitem{LiF2020} J. -K.  Xie,  S.-L.  Ma,  and F.-L.  Li,  Phys.  Rev.  A \textbf{101}, 042331 (2020).
\bibitem{Li2018} J. Li, S.-Y. Zhu, and G. S. Agarwal, Phys. Rev. Lett. \textbf{121}, 203601 (2018).
\bibitem{Li2020} M. Yu, H. Shen, and J. Li, Phys. Rev. Lett. \textbf{124}, 213604 (2020).
\bibitem{Yuan2020} H. Y. Yuan , P. Yan, S. Zheng, Q. Y. He, K. Xia, and M.-H. Yung, Phys. Rev. Lett. \textbf{124}, 053602 (2020).
\bibitem{Yuan2020-PRB} H. Y. Yuan , S. Zheng, Z. Ficek, Q. Y. He,  and M.-H. Yung, Phys. Rev. B \textbf{101}, 014419 (2020).
\bibitem{Qian2021} D.-W. Luo,  X.-F.  Qian,  and T.  Yu,  Opt.  Lett. \textbf{46},  1073 (2021).
\bibitem{Sharma2021} S.  Sharma, V. A. S. V.  Bittencourt,  A. D.  Karenowska,  and S. V.  Kusminskiy,  Phys.  Rev.  B \textbf{103}, L100403 (2021).
\bibitem{QYHe2021} F.-X.  Sun, S.-S.  Zheng,  Y.  Xiao,  Q.  Gong,  Q.  He,  and K. Xia,  Phys.  Rev.  Lett.  \textbf{127}, 087203 (2021).
\bibitem{Ruoso2017} C. Braggio, G. Carugno, M. Guarise, A. Ortolan, and G. Ruoso, Phys. Rev. Lett. \textbf{118}, 107205 (2017).
\bibitem{Weides2020} T. Wolz, A. Stehli, A. Schneider, I. Boventer, R. Mac\^{e}do, A. V. Ustinov, M. Kl\"{a}ui, and M. Weides, Commun. Phys. \textbf{3},3 (2020).
\bibitem{GlauberPRL} R. J.  Glauber,  Phys. Rev. Lett.  \textbf{10}, 84 (1963).
\bibitem{Glauber1963-1}  R. J.  Glauber, Phys. Rev.  \textbf{130}, 2529(1963).
\bibitem{Glauber1963-2}  R. J.  Glauber, Phys. Rev.  \textbf{131}, 2766(1963).
\bibitem{HBT} R.  Hanbury Brown and R. Q.  Twiss,  Nature,  \textbf{177},  27 (1956). 
\bibitem{QuanOpt1} C.  Gerry and P.  Knight,  \emph{Introductory Quantum Optics} (Cambridge University Press,  Cambridge,  England,  2004).
\bibitem{QuanOpt2} M. O.  Scully and M.  Suhail Zubairy,  \emph{Quantum Optics} (Cambridge University Press,  Cambridge,  England,  1997).
\bibitem{QuanOpt3}  D. F.  Walls and G.J.  Milburn,  \emph{Quantum Optics} (Springer-Verlag Berlin Heidelberg,  2008).
\bibitem{Rebic2009} S.  Rebi\'{c}, J.  Twamley,  and G. J.  Milburn,  Phys.  Rev.  Lett.  \textbf{103}, 150503 (2009).
\bibitem{Bozyigit2011} D.  Bozyigit,  C.  Lang,  L.  Steffen,  J. M.  Fink,  C.  Eichler,  M.  Baur,  R.  Bianchetti,  P. J.  Leek,  S. Filipp,  M. P. da Silva,  A. Blais,  and A.Wallraff,  Nat. Phys.  \textbf{7},  154 (2011).
\bibitem{LangC2011} C.  Lang,  D.  Bozyigit,  C.  Eichler,  L.  Steffen,  J. M.  Fink,  A. A.  Abdumalikov,  Jr.,  M. Baur,  S. Filipp,  M. P. da Silva,  A. Blais,  and A. Wallraff,  Phys.  Rev.  Lett. \textbf{106},  243601 (2011).
\bibitem{EichlerC2012} C.  Eichler,  D. Bozyigit,  and A.  Wallraff,  Phys.  Rev.  A \textbf{86},  032106 (2012).
\bibitem{LangC2013} C.  Lang,  C.  Eichler,  L.  Steffen,  J. M.  Fink,  M. J.  Woolley,  A.  Blais,  and A.  Wallraff,  Nat.  Phys.  \textbf{9},  345 (2013).
\bibitem{Peng2016} Z.H. Peng,  S.E. de Graaf,  J.S. Tsai,  and O.V. Astafiev,  Nat. Commun.  \textbf{7},  12588 (2016). 
\bibitem{Gasparinetti2017} S.  Gasparinetti,  M.  Pechal,  J.-C.  Besse,  M.  Mondal,  C.  Eichler,  and A.  Wallraff,  Phys.  Rev.  Lett.  \textbf{119},  140504(2017).
\bibitem{Rolland2019} C. Rolland,  A. Peugeot,  S. Dambach,  M. Westig,  B. Kubala, Y. Mukharsky,  C. Altimiras,  H. le Sueur,  P. Joyez,  D. Vion,  P. Roche,  D. Esteve,  J. Ankerhold,  and F.  Portier,  Phys.  Rev.  Lett.  \textbf{122},  186804 (2019).  
\bibitem{Higher1} M. A$\beta$mann,  F.  Veit,  M.  Bayer,  M. van der Poel,  and J. M.  Hvam,  Science \textbf{325},  297 (2009).  
\bibitem{PRL2018} M. Klaas,  H. Flayac,  M. Amthor,  I. G. Savenko,  S. Brodbeck,  T.  Ala-Nissila,  S.  Klembt,  C.  Schneider,  and S. H\"{o}fling,  Phys.  Rev.  Lett.  \textbf{120},  017401 (2018).
\bibitem{NatMat2019} G.  Mu\~{n}oz-Matutano,  A.  Wood,  M.  Johnsson,  X.  Vidal,  B. Q.  Baragiola,  A.  Reinhard,  A.  Lema\^{i}tre ,  J.  Bloch,  A.  Amo,  G.  Nogues,  B.  Besga,  M.  Richard,  and T.  Volz,  Nat.  Mater.  \textbf{18},  213(2019).
\bibitem{OptoMech1} P. Rabl,  Phys.  Rev.  Lett.  \textbf{107}, 063601 (2011).
\bibitem{OptoMech2} A. Nunnenkamp, K. B\o rkje, and S. M. Girvin,  Phys.  Rev.  Lett.  \textbf{107}, 063602 (2011).
\bibitem{Atom1} M.  Yasuda and F.  Shimizu,  Phys.  Rev.  Lett.  \textbf{77},  3090 (1996).
\bibitem{Atom2} S. Folling, F. Gerbier, A. Widera, O. Mandel, T. Gericke, and I. Bloch,  Nature \textbf{434},  481 (2005).
\bibitem{Atom3} M.  Schellekens,  R. Hoppeler,  A. Perrin,  J. Viana Gomes,  D. Boiron,  A. Aspect,  and  C. I.  Westbrook,  Science \textbf{310}, 648 (2005).
\bibitem{Atom4} A.  \"{O}ttl,  S.  Ritter,  M.  K\'{o}hl,  and T.  Esslinger,  Phys.  Rev.  Lett.  \textbf{95},  090404(2005).  
\bibitem{Atom5} T. Rom,  T. Best,  D. van Oosten,  U. Schnieder,  S. Folling,  B. Paredes,  and I. Bloch,   Nature \textbf{444}, 733 (2006).
\bibitem{Atom6} T. Jeltes,  J. M. McNamara,  W. Hogervorst,  W. Vassen,  V. Krachmalnicoff,  M. Schellekens,  A. Perrin,  H. Chang,  D. Boiron,  A. Aspect,  and C. I. Westbrook,  Nature \textbf{445}, 402 (2007).
\bibitem{Higher3} S. S.  Hodgman,  R. G. Dall,  A. G. Manning,  K. G. H. Baldwin,  and A. G. Truscott,  Science  \textbf{331},  1046 (2011).
\bibitem{Atom7} R. G.  Dall,  A. G.  Manning,  S. S.  Hodgman,  W.  Rugway,  K. V.  Kheruntsyan,  and A. G.  Truscott,  Nat. Phys. \textbf{9}, 341(2013).
\bibitem{Atom8} P. M. Preiss, J. H. Becher, R. Klemt, V. Klinkhamer,  A. Bergschneider,  N.  Defenu,  and S. Jochim,  Phys. Rev. Lett.  \textbf{122}, 143602(2019).
\bibitem{Atom9} H.  Cayla,  S.  Butera ,  C.  Carcy,  A.  Tenart,  G.  Herc\'{e},  M.  Mancini,  A.  Aspect ,  I.  Carusotto,  and D.  Cl\'{e}ment,  Phys.  Rev.  Lett.  \textbf{125}, 165301 (2020).
\bibitem{SpinCurr} S. A.  Bender,  A.  Kamra,  W.  Belzig,  and R. A.  Duine,  Phys.  Rev.  Lett.  \textbf{122}, 187701 (2019).
\bibitem{Hu2019PRB} C.  Match,  Michael Harder,  Lihui Bai,  Paul Hyde,  and Can-Ming Hu,  Phys.  Rev.  B \textbf{99}, 134445 (2019).
\bibitem{Leggett1983} A.O. Caldeira and A. Leggett, Physica A \textbf{121}, 587 (1983).
\bibitem{daSilva2010} M. P.  da Silva,  D.  Bozyigit,  A.  Wallraff,  and A.  Blais, Phys. Rev.  A \textbf{82},  043804 (2010).
\end{thebibliography}
\end{document}